\documentclass[amssymb,amsmath,aps,secnumarabic,floatfix,prl,showpacs]{revtex4}
\usepackage{pstricks,pst-node,graphics}
\usepackage{psfig}
\usepackage{epsfig}
\expandafter\ifx\csname package@font\endcsname\relax\else
 \expandafter\expandafter
 \expandafter\usepackage
 \expandafter\expandafter
 \expandafter{\csname package@font\endcsname}%
\fi

\begin{document}
\title{The Oslo rice pile model is a quenched Edwards-Wilkinson equation}
\author{Gunnar Pruessner}
\email{gunnar.pruessner@physics.org}
\affiliation{
Department of Mathematics,
Imperial College,
180 Queen's Gate,
London SW7 2BZ,
UK}
\date{\today}

\begin{abstract}
The Oslo rice pile model is a sandpile-like paradigmatic model of
 ``Self-Organized Criticality'' (SOC).  In this paper it is shown that
 the Oslo model is in fact \emph{exactly} a discrete realization of the
 much studied quenched Edwards-Wilkinson equation (qEW)
 {\bracketOpen}Nattermann et~al., J. Phys. II France \textbf{2}, 1483
 (1992){\bracketClose}. This is possible by choosing the correct
 dynamical variable and identifying its equation of motion. It
 establishes for the first time an exact link between SOC models and the
 field of interface growth with quenched disorder. This connection is
 obviously very encouraging as it suggests that established theoretical
 techniques can be brought to bear with full strength on some of the
 hitherto elusive problems of SOC.
\end{abstract}

\maketitle

The Oslo rice pile model (Oslo model hereafter) was originally intended
to model the relaxation processes in real rice piles
\cite{FretteETAL:1996}. Meanwhile, it has been subject to many
investigations and publications in its own right. The model as defined
below supposedly develops into a scale free state without the explicit
tuning of external parameters, and is therefore regarded as an example
of Self-Organized Criticality (SOC) \cite{Jensen:98}. In fact, contrary
to many other ``standard'' models of SOC
\cite{DattaChristensenJensen:2000,Dorn:2001,LisePaczuski:2001b,JensenPruessner:2002b},
it shows a reliable and consistent (simple) scaling behavior and is
robust against certain changes in the details of the dynamics
\cite{Zhang:1997,BengrineETAL:1999,BengrineETAL:1999b}. The most
prominent observable in the model, the avalanche size $s$, is governed
by a probability distribution $\PC(s)$ which obeys simple scaling,
\begin{equation}
 \PC(s) = s^{-\tau} \GC(s/s_0) \text{ and } s_0 = L^D \text{,}
\label{eq:def_tau}
\end{equation}
where $L$ denotes the system size and $\tau$ and $D$ are critical
exponents, consistently reported to be $\tau=1.55(10)$ and $D=2.25(10)$
\cite{ChristensenETAL:1996,PaczuskiBoettcher:1996,Zhang:1997,BengrineETAL:1999,CorralPaczuski:1999,BengrineETAL:1999b}.
These two exponents are related by $D(2-\tau)=1$
\cite{ChristensenETAL:1996,PaczuskiBoettcher:1996}, which can be proven
easily given that the first moment of $\PC(s)$, $\ave{s}$, scales like $L$. 

In the following the model is defined, the relevant dynamical variable
extracted and its equation of motion derived, which turns out to be a
discretized quenched Edwards-Wilkinson (qEW) equation.
By analyzing the essential characteristics of the model on the
lattice, such as uniqueness of the solution and symmetries, it is
then possible to construct the continuum theory,
which can subsequently be examined using standard methods.

The model \cite{ChristensenETAL:1996} is defined on a one dimensional
grid of size $L$, where each site $i=1 \cdots L$ has slope $z_i$ and
critical slope $z^c_i \in \{1,2\}$. Starting from an initial
configuration with $z_i=0$ and $z^c_i$ random everywhere, the model
evolves according to the following update rules: 1) (Driving) Increase
$z_1$ by one. 2) (Toppling) If there is an $i$ with $z_i > z^c_i$
decrease $z_i$ by $2$ and increase its nearest neighbors by one,
$z_{i\pm 1} \to z_{i\pm 1} + 1$, provided that $1\le i\pm 1 \le L$. A
new $z^c_i$ is chosen at random, $1$ with probability $p$ and $2$ with
probability $q\equiv 1-p$. 3) Repeat the second step until $z_i \le
z^c_i$ everywhere.  Then proceed with the first step. The order of
updates is irrelevant in this model and the original definition does not
fix it explicitly. Therefore the microscopic (fast) timescale is
\emph{a priori} undefined.

The avalanche size $s$ is defined as the number of charges, i.e. apart
from boundary effects, twice the number of times the second rule is
applied between two consecutive application of the first rule. For
convenience the model is dissipative on both boundaries, where one of
the two ``units'' lost by the boundary site during toppling leaves the
system. 

A few years ago Paczuski and Boettcher translated the Oslo model into
the language of interfaces in random media
\cite{PaczuskiBoettcher:1996}. However, the evolution of the dynamical
variable $H(x,t)$, which is the total number of topplings of site $x$,
was given by $\partial_t H = \theta ( \laplace H - \eta(x, H))$, where
$\partial_t$ is defined in discrete time, i.e. $\partial_t H \equiv
H(x,t+1)-H(x,t)$ and $\laplace$ is the lattice Laplacian, so that $x$ is
actually an index. The last term $\eta(x,H)$ represents a quenched
noise. The Heaviside $\theta$-function makes this equation of motion
highly nonlinear and analytically almost intractable
\cite{Diaz-Guilera:1992}.  Paczuski and Boettcher have already conjectured
that the Oslo model is in the same universality class as qEW
\cite{NattermannETAL:1992}. More recently, Alava has suggested that
certain other sandpile models are described by qEW
\cite{Alava:2002}. It is, however, important to realize that no rigorous
and exact link has so far been established between SOC models and the
qEW equation.

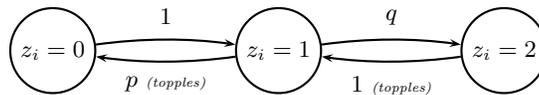
\begin{figure}
\begin{center}
\begin{pspicture}(0,0)(8,1.7)
\rput(1,1){\circlenode{A}{$z_i=0$}}
\rput(4,1){\circlenode{B}{$z_i=1$}}
\rput(7,1){\circlenode{C}{$z_i=2$}}
\ncarc{->}{A}{B}\naput{$1$}
\ncarc{->}{B}{C}\naput{$q$}
\ncarc{->}{C}{B}\naput{$1$ \emph{\tiny{(topples)}}}
\ncarc{->}{B}{A}\naput{$p$ \emph{\tiny{(topples)}}}
\end{pspicture}
\end{center}
\caption{Each site can be in one of three states and changes stepwise
 between them, whenever it receives a charge. The labels indicate the
 probability of the move and whether it entails a toppling. \label{fig:three_states}}
\end{figure}
The crucial step to make this correspondence exact is to identify the
proper dynamical variable. It is found in the form of the number of
times a site has been charged 
(i.e. received a unit from a neighbor during a toppling or by external
drive, see below)
$h(x,t)$, where $x$ and $t$ are discrete for the
time being. There is a simple functional relation
between $h(x,t)$ and $H(x,t)$, which can be obtained as follows: Each
site can be in one of three stable configurations, $z_i \in {0,
1, 2}$. When a site receives a unit from a neighbor, it changes state
as shown in Fig.~\ref{fig:three_states}. Charging a site in state $0$
necessarily leads to state $1$ without toppling and the specific value
of $z_i^c$ is completely irrelevant at this stage. Similar for
state $2$: If a site receives a charge in this state, its $z_i^c$
must be $2$ and it must topple. The only point where the value of
$z_i^c$ actually matters, is in state $1$, therefore it can be
effectively chosen at random when necessary, so that the site topples with
probability $p$ (according to the probability of having $z_i^c=1$) or
increases to $2$ with probability $q$ (see
Fig.~\ref{fig:three_states}). It is immediately clear that any even
number of charges, say $m=2n$, starting from $z_i=1$ leads to state $1$ 
again with $n$ topplings. An odd number of charges, say $m=2n+1$, leads
either to $n$ topplings and state $2$ or $n+1$ topplings and state
$0$. This is illustrated in Fig.~\ref{fig:three_states}: The $m$ charges
lead to $m$ steps along the arrows. Whenever one moves left, the
site topples.

In order to write a functional relation between $h(x,t)$ and $H(x,t)$,
the randomness in the decision of moving to the left or to the 
right from state $1$ must be quenched in $h(x,t)$, i.e. it is not
allowed to change unless $h(x,t)$ changes. This can be summarized as
\begin{equation}
 H(x, t+1) = \half \left( h(x,t) + \eta(x, h(x,t)) \right) \ \text{,}
\label{eq:h_to_H}
\end{equation}
where $\eta$ is $0$ whenever $h(x,t)$ is even, corresponding to state
$1$. If $h(x,t)$ is odd, $\eta$ is either $1$ (with probability $p$,
state $z_i=0$) or $-1$ ($z_i=2$). 
Every sequence of $\eta(x,h)$ values maps uniquely to a sequence of
$z_i^c$ and vice versa. 
The equation above can easily
be transformed to comply to any initial configuration, especially to 
$z_i(t=0)\equiv 0$. Essentially, it is 
(\ref{eq:h_to_H}),
which makes the exact identification of the Oslo model
and qEW possible. 

The final equation is derived by noting that obviously $h(x,t) =
H(x-1,t) + H(x+1,t)$ with appropriately chosen boundary conditions (BC's)
(see below), so that using the short hand notation 
$h^\pm = h(x\pm 1,t)$ and
$\eta^\pm = \eta(x\pm 1,h^\pm)$ the equation of motion is
\begin{equation}
\label{eq:oslo_qEW_discrete}
 h(x,t+1) - h(x,t) =  \half \left( h^-\!\! - 2 h(x,t) + h^+ 
+
\eta^+\!\! + \eta^- \right) \ \text{,}
\end{equation}
which is the \emph{exact} representation of the Oslo model as defined
above, captured in a single equation. Its differential form is accordingly
\begin{equation}
 \partial_t h(x,t) = \half \laplace h(x,t) + 
 \left( 1 + \half \frac{d^2}{dx^2} \right) \eta(x,h(x,t)) \ \text{.}
\label{eq:oslo_qEW_1}
\end{equation}
The right hand BC is $h(x=L+1,t)\equiv 0$ (and
$h(x=L,t)\equiv 0$ in the continuum),
while the left hand BC provides the driving via $h(x=0, t)= 2 E(t)$,
$E(t)$ being the total number of initial seeds (step 1 above) at time 
$t$. These seeds arrive at site $x=1$ via the Laplacian. In the
continuum, the simplest drive is $E(t)=v t$ with $v$ a driving
velocity and $t$ the microscopic time. Together with the BC's,
Eq.~(\ref{eq:oslo_qEW_1}) or the generalized form 
\begin{equation}
 \partial_t h(x,t) = \kappa \laplace h(x,t) + 
 g \left( 1 + \lambda \frac{d^2}{dx^2} \right) \eta(x,h(x,t)) \ \text{,}
\label{eq:oslo_qEW}
\end{equation}
where the correlator of $\eta$ is now normalized, i.e.  $\int dx \int dh
\ave{\eta \eta} = 1$, describes the movement of an elastic band over a
rough surface \cite{DongETAL:1993} pulled by a transverse force acting at one end
point only. Below it is shown that the $\lambda$-term disappears in the
continuum, establishing the \emph{first rigorous identification} of the
Oslo model and the qEW equation.  The same equation with different
properties of 
the noise term and/or different BC's applies to other models, such as the BTW
model \cite{BakTangWiesenfeld:1987}, Fixed Energy Sandpiles (for example
\cite{Vespignani:2000}) or the tilted sandpile
\cite{Malthe-Sorensen:1999}. 
Having identified the relevant dynamical variable $h$, 
the effect of modifications of the
dynamical rules of the Oslo model, such as
\cite{Zhang:1997,BengrineETAL:1999,BengrineETAL:1999b}, can be
understood.

The equation above exemplifies a general ``trick''\footnote{The
substitution was already used earlier (see \cite{Alava:2002} and
references therein).}  to get rid of $\theta$-functions in equations of motion
--- they often appear in descriptions of sandpile-like systems (for
example \cite{Diaz-Guilera:1992}): One simply replaces $\theta(h-h_c)$
by $h+\eta(h)$ with an appropriately chosen sawtooth-like $\eta$. This
does not necessarily simplify the problem, unless there is already a
quenched noise present in the system. In this case the $\theta$ turns
into a correlation in $\eta$. This is highly remarkable from the point
of view of SOC, because the presence of ``thresholds'' is usually
expected to be a crucial ingredient of SOC
\cite{BakTangWiesenfeld:1987,CalfieroETAL:1995,Jensen:98}. Moreover, the
correlations in $\eta$, which are of fundamental significance in
interface models \cite{NattermannETAL:1992,LeschhornETAL:1997} and have
been neglected in former mappings, now arise naturally from the
dynamical description of the model.

In order to construct the proper continuum theory, it is worthwhile to
consider the formal solution of Eq.~(\ref{eq:oslo_qEW}). It will turn out
later that $E(t) = v t$ is sufficiently general, so that it makes sense
to define $v(x) \equiv v \frac{L-x}{L}$ and
\begin{equation}
 h(x,t) = 2 v(x) t + P_3(x) + z(x,t)
\label{eq:data_shift}
\end{equation}
in order to homogenize the BC's. $P_3(x)$ is a third order polynomial
only present to cancel the first term in the differential equation,
i.e. $\kappa \partial_x^2 P_3 = 2 v(x)$, with roots at $x=0$ and
$x=L$. Therefore $\partial_t z = \kappa \laplace z + g \eta_\lambda(x,
h(x,t))$ with homogenous BC's. The term $ \eta_\lambda(x, h(x,t)))
\equiv ( 1 + \lambda \frac{d^2}{dx^2} ) \eta(x,h(x,t))$ is actually a
functional of $h$. The initial condition of $z(x,t)$ is not
$z(x,t=0)\equiv 0$ as for $h$, because of the data shift above. But due
to the homogenous BC's any initial condition decays, so that the initial
sources, accounting for $z(x,t=0)=-P_3(x)$, can be ignored. Then the
formal solution is $ z(x,t) = \sum_{n=1}^\infty z_n(t) \sin(k_n x)$ with
\begin{eqnarray}
 z_n(t) &=& \frac{2 g}{L} \int_0^t dt' \int_0^L dx' 
\eta_\lambda(x', 2 v(x) t + z(x',t)) \nonumber \\
& \times & \sin(k_n x)\ \exp(-k_n^2 \kappa (t-t')) 
\label{eq:znt}
\end{eqnarray}
and $k_n=\frac{\pi n}{L}$.

According to Eq.~(\ref{eq:data_shift}), the tilt of $h(x,t)$ in $x$
increases in time. Assuming stationarity of the relevant statistical
properties (especially avalanches as defined below), this requires the solution
to be invariant under tilt, which is also known as Galilean invariance
\cite{Meakin:1998}: $h' = h+ \alpha x$ must produce the same statistics
as $h$, which entails $\eta(x, a + \alpha x) $ to be equally likely as
$\eta(x, a)$, so that
$ 
\ave{ \eta(x, a + \alpha x) \eta(x', a' + \alpha x' } = \ave{ \eta(x, a ) \eta(x', a' ) } 
$.  
But assuming the standard form \cite{NattermannETAL:1992} 
$\ave{ \eta(x, a ) \eta(x', a' ) } = \Delta_{\parallel}(x-x') \Delta_{\perp}(a-a')$,
the correlator obeys for any $x-x'$ where $\Delta_{\parallel}(x-x')$ is
finite,
$\Delta_{\perp}(a-a') =\Delta_{\perp}(a-a'+\alpha(x-x'))$.
This holds for any $\alpha$, so if
$\Delta_{\parallel}(x-x')$ was finite for any $x-x'\ne 0$,
$\Delta_{\perp}$ would be bound to be a constant. This is impossible,
because $\Delta_{\perp}$ must be non-vanishing somewhere and
normalizable, so that $\Delta_{\parallel}(x-x')$ must vanish for any
finite $x-x'$, i.e. it
\emph{must} be a $\delta$-function.

Next it can be shown that the Oslo model obeys Middleton's
no-passing \cite{Middleton:1992}. For $\lambda \ne 0$ this will lead to
a constraint on the noise which is incompatible with the $\delta$
correlation of $\Delta_{\parallel}$ in the
continuum, so that $\lambda$ must vanish in the continuum. Defining a
partial ordering $\succeq$ for two configurations $h_1(t_1, x)$ and
$h_2(t_2, x)$ of the interfaces as 
$
 h_1(t_1, x) \succeq h_2(t_2, x) \Leftrightarrow
\forall_{x\in [0,L]}
h_1(t_1, x) \ge h_2(t_2, x) 
$, 
one has to show that this order is preserved under the
dynamics \cite{SethnaDahmen_etal:1993}. With the ``external field''
being the BC's $E_1(t)$ and $E_2(t)$, one shows that if
$h_1(t_0, x) \succeq h_2(t_0, x)$ for a given $t_0$ (which entails
$E_1(t_0)\ge E_2(t_0)$) the interfaces can never ``overtake'' each other
at $t\ge t_0$. By assuming the opposite,
one only needs to prove that where the two interfaces ``touch'' for the first
time, $x_0$, the velocity of $h_1$ is higher or equal to the
velocity of $h_2$. For the model on the lattice
(\ref{eq:oslo_qEW_discrete}), this is equivalent to
\begin{equation}
h_1^+ + \eta_1^+
+ h_1^- + \eta_1^-
\ge
h_2^+ + \eta_2^+
+ h_2^- + \eta_2^-
\label{eq:no-passing_condition_discrete}
\end{equation}
using the same notation as in (\ref{eq:oslo_qEW_discrete}).  In the
original discrete model, condition (8) follows immediately from
$\eta(x,h)+h$ being a monotonically increasing function in $h$ for any
$x$.  For the continuum equation (\ref{eq:oslo_qEW}) the corresponding
calculation gives
\begin{equation}
 \lambda g \partial_h \eta(x, h) \ge - \kappa
\label{eq:no-passing_condition}
\end{equation}
assuming that 
$
\frac{d^2}{dx^2} \eta =
\partial_x^2 \eta +
\partial_x h \partial_x \partial_h \eta +
\partial_x h \partial_h \partial_x \eta +
\partial_x^2 h \partial_h \eta +
(\partial_x h)^2 \partial_h^2 \eta
$
and that the interface is smooth in $x_0$ such that 
$\partial_x h_1(x_0,t) = \partial_x h_2(x_0,t)$ and 
$\partial_x^2 h_1(x_0,t) > \partial_x^2 h_2(x_0,t)$. 
For a noise with divergent width, $\Delta_{\parallel}(x) = \delta(x)$,
Eq.~(\ref{eq:no-passing_condition}) cannot hold for any $\lambda\ne 0$,
i.e. a non-vanishing $\lambda$ destroys no-passing. However, no-passing
must be regarded as a crucial feature, as it ensures the asymptotic
uniqueness of the configuration and is reminiscent of the irrelevance of
the order of updates in the original model, so that $\lambda =0$ is a
necessary condition for the equivalence of the continuum and discrete
model. 

This is physically justified: Assuming a smooth $\eta$, in the continuum
approximation of Eq.~(\ref{eq:oslo_qEW_discrete}) $\lambda$ becomes
proportional to the square of the lattice spacing and therefore vanishes
in the continuum limit.

Keeping the $\lambda$ term nevertheless, a na\"{\i}ve scaling analysis
shows that it is irrelavant. Moreover, its Fourier transform in
Eq.~(\ref{eq:znt}) produces only a term $-g\lambda k_n^2$, because of the
total derivative in $\eta_{\lambda}$. This can be absorbed into the bare
propagator of a perturbative expansion in the style of
\cite{NattermannETAL:1992,LeschhornETAL:1997} in the form
$
 \frac{2g (1-\lambda k_n^2) }{L (\kappa k_n^2 + i \omega)}
$,
leading possibly to an ultraviolet divergence. Apart from that, the
terms obtained for an renormalization group treatment are structurally
the same as in 
\cite{LeschhornETAL:1997} as calculations show (details to be published
later). The only differences are due to the peculiar way of driving the
interface (i.e. the term $2v(x)$, which is a mean velocity in
(\ref{eq:data_shift}), but also drives the model by moving the quenched
noise in (\ref{eq:znt})) and the non-conservative nature of the
interface 
(which makes sense only for a finite system) 
leading to the
homogenous BC's and therefore to the $\sin(k_n x)$ rather than $\exp(2 i
k_n x)$ terms.  In turn, the standard qEW problem
\cite{NattermannETAL:1992} corresponds to an Oslo
model with periodic BC's 
and continuous, uniform drive.

Expanding $\eta$ in powers of $z_n$, the first two terms of
$z_n(\omega)$ (the Fourier transform of (\ref{eq:znt}) in $t$) are: 
\begin{eqnarray}
 z_n(\omega) &=& \frac{2 g (1-\lambda k_n^2)}{L (\kappa k_n^2 + i \omega)}
\Big( \int_0^L dx' \hat{\eta}\left(x', \frac{\omega}{2v(x')}\right) 
\frac{\sin(k_n x')}{2 v(x')}  \nonumber \\
&+&     \int_0^L dx' \int_{-\infty}^\infty dq \sum_{m=1}^\infty \hat{\eta}(x',q) 
\frac{i q \sin(k_m x')}{\sqrt{2\pi}} z_m(\omega - 2 v(x') q) \sin(k_n x') \Big) \nonumber 
\end{eqnarray}
where $\hat{\eta}(x,q)$ is the Fourier transform of $\eta(x,h)$ in
$h$. 

The definition of the avalanche size $s$ in the continuum is the area
between the interface configurations at two times $t_1$ and $t_2$, 
$
 s=\int_0^L dx (h(x,t_2)-h(x,t_1))
$,
so that $\ave{s}= v \Delta t L$ with $\Delta t \equiv t_2-t_1$, because
$\ave{z(x,t)}$ is expected to be 
asymptotically independent of $t$, as a
non-vanishing $\lim_{t\to \infty}\partial \ave{z(x,t)}$ 
with homogenous BC's would require
support for a divergent
curvature of the interface. Choosing $\Delta h \equiv  \Delta t
v$ constant for different system sizes $L$ then preserves the
property $\ave{s} \propto L$. 

Due to the asymptotic uniqueness of the solution the system can either
be driven in jumps of $\Delta h$ separated by sufficiently long times,
or driven very slowly taking ``snapshots'' of the configuration in order
to calculate $s$.

The model possesses two characteristic timescales: One is the diffusive
timescale $t_0 \equiv L^2/\kappa$, the other one is the non-trivial
scale due to noise and drive, $t_g \equiv g^2/(v^3 L)$. One has to
maintain a sufficiently large $\Delta t$ to prevent distinct avalanches
from merging, otherwise the central limit theorem would turn $\PC(s)$
into a Gaussian. The SOC limit is usually identified with $v\to 0$,
which makes sense only in the presence of an intrinsic scale for $v$.
The only combination of parameters ($\kappa$, $g$ and $L$, but
$\lambda=0$) which provides a ``natural velocity'' is $v_g \equiv (g^2
\kappa)^{1/3}/L$. The SOC condition $v\to 0$ is therefore already met by
$v \ll v_g \propto L^{-1}$, which is however, not sufficient. 
According
to Ref.~\cite{PaczuskiBoettcher:1996} $\Delta t \gg L^z$ with $z\approx
1.42$,
so that $\Delta h=\text{const.}$ entails $v \ll L^{-z}$, which
therefore seems to be the correct condition for SOC,
even though the microscopic timestep in \cite{PaczuskiBoettcher:1996} is defined as a parallel
      update, which is not \emph{exactly} (\ref{eq:oslo_qEW_discrete}).

\begin{figure}[th]
\includegraphics[width=0.9\linewidth]{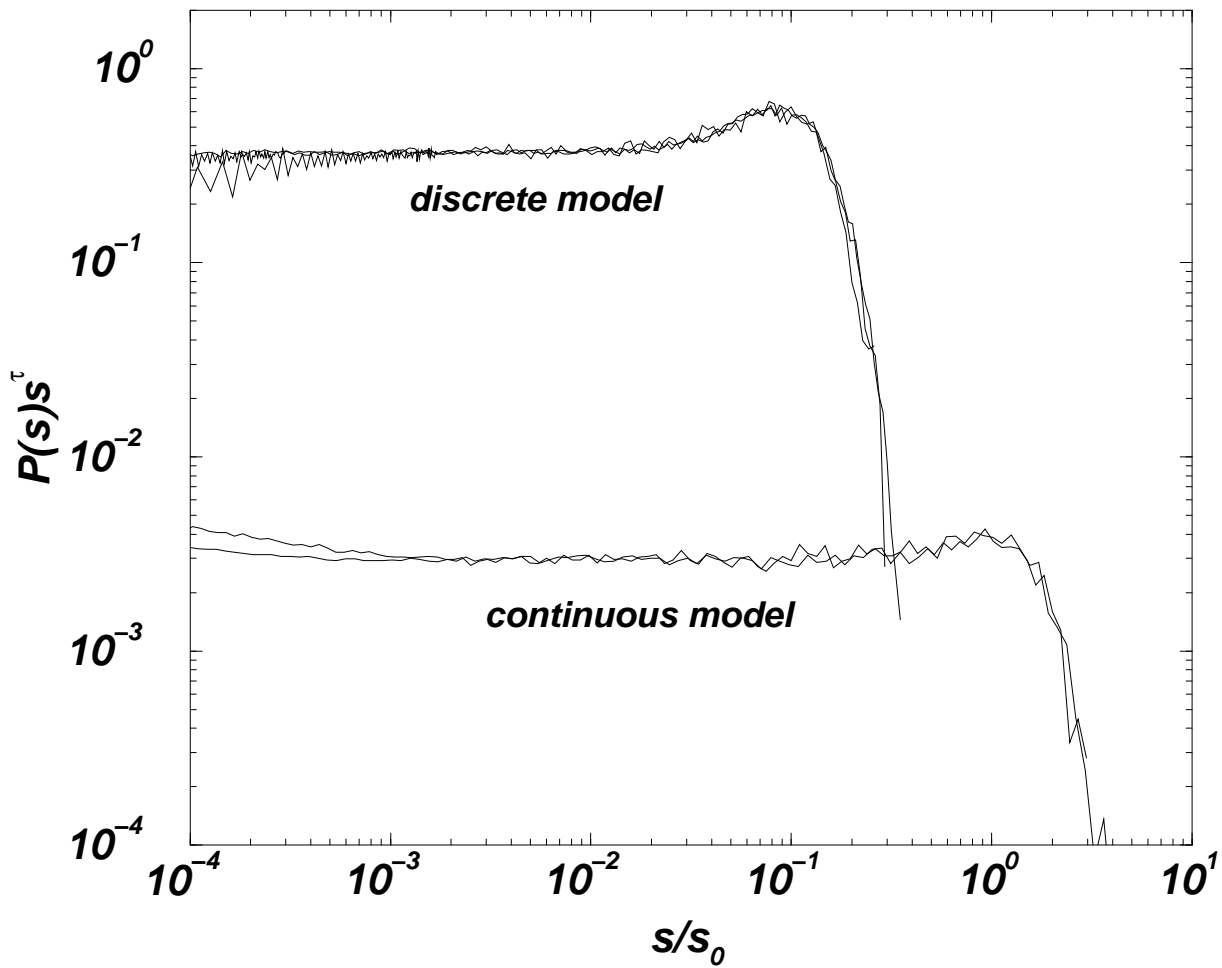}
\caption{Comparison of a data collapse according to (\ref{eq:def_tau})
for system sizes between $L=128$ and $L=512$ and the continuous and the
discrete realization of the model.  The same value of $\tau=1.55$
collapses all curves within each model onto its scaling function.  Due
to the omission of the non-universal constants in Eq.~(\ref{eq:def_tau}) the
two resulting curves are shifted relative to each other.
  \label{fig:data_example}}
\end{figure}

Preliminary numerical studies indeed suggest that
(\ref{eq:oslo_qEW}) with $\lambda=0$ is a valid continuous
description of the Oslo model: Fig.~\ref{fig:data_example} compares a
scaling collapse for different system sizes of the continuous model
(with $\lambda=0$) and the original, discrete one. The best collapse is
obtained by $\tau=1.55$ for both models.
 The scaling law $D=1+\chi$ \cite{PaczuskiBoettcher:1996} remains applicable
as long as the two configurations at $t_1$ and $t_2$ are
correlated. It is in perfect agreement with numerical results
\cite{Leschhorn:1993,LeDoussalWieseChauve:2002} for the qEW model
\footnote{For $\chi = 5/4$ one has $D=9/4$ and $\tau=14/9$.}.

In conclusion, the Oslo model has been reduced to a quenched
Edwards-Wilkinson equation. In the continuum limit the qEW becomes the
\emph{exact} equation of motion for the Oslo model. This not only makes
it possible to approach the exponents of an SOC-model analytically, but
also gives insight into the nature of avalanche like behavior and the
relation between SOC and other theories of critical phenomena. It
provides the perfect test bed for analytical methods proposed for SOC.

The established relationship is presently being pursued in order to
develop a direct approach to the critical exponent $\tau$, clear up the
r\^{o}le of the noise and clarify the relation between noise and
drive. The framework used here is also promising for other models, such
as the BTW model \cite{BakTangWiesenfeld:1987}, various other sandpile models 
\cite{Vespignani:2000,Malthe-Sorensen:1999} and the Zhang model
\cite{Zhang:1989}. 

\begin{acknowledgments}
The author wishes to thank Nicholas R. Moloney for proof reading, Henrik
 J. Jensen for suggesting the problem and for very helpful discussions,
 as well as Alvin Chua and Kim Christensen for presenting and discussing
 their work \cite{ChuaChristensen:2002} prior to publication.  The
 author gratefully acknowledges the support of the EPSRC and the NSF during
 the 2001 Boulder School for Condensed Matter and Material Physics.
\end{acknowledgments}
\bibliography{articles,books}
\end{document}